\documentclass[draft]{agujournal2019}
\usepackage{url} %this package should fix any errors with URLs in refs.
\usepackage{lineno}
\usepackage{xr}
\usepackage{pdfpages}
\usepackage[inline]{trackchanges} %for better track changes. finalnew option will compile document with changes incorporated.
\usepackage{soul}
\usepackage{natbib}
\usepackage{amsmath}
\draftfalse

\journalname{Geophysical Research Letters}

\begin{document}

%%%%%%%%%%%%%%%%%%%%%%%%%%%%%%%%%%%%%%%%%%%%%%%
%  TITLE
%
% (A title should be specific, informative, and brief. Use
% abbreviations only if they are defined in the abstract. Titles that
% start with general keywords then specific terms are optimized in
% searches)
%
%%%%%%%%%%%%%%%%%%%%%%%%%%%%%%%%%%%%%%%%%%%%%%%

% Example: \title{This is a test title}

\title{Quantifying the Influence of Climate on Storm Activity Using Machine Learning}

\authors{Or Hadas\affil{1}, Yohai Kaspi\affil{1}}

\affiliation{1}{Department of Earth and Planetary Sciences, Weizmann Institute of Science, Rehovot, Israel}

\correspondingauthor{Or Hadas}{or.hadas@weizmann.ac.il}

\begin{keypoints}
\item Using 84 years of ERA-5 reanalysis and thousands of storm tracks, a machine-learning approach quantifies how climate shapes storm activity
\item While the climate strongly influences climatological storm activity, individual storms remain dominated by synoptic variability
\item Thus, attribution efforts should focus on a climatic perspective or focusing on factors directly connected to anthropogenic climate change
\end{keypoints}

\begin{abstract}
Extratropical storms shape midlatitude weather and vary due to the slowly evolving climate and the rapid changes in synoptic conditions. While the influence of each factor has been studied extensively, their relative importance remains unclear. Here, we quantify the climate's relative importance in mean storm activity and individual storm development using 84 years of ERA-5 data and convolutional neural networks. We find that the constructed model predicts over 90\% of the variability in the mean storm activity. However, a similar model predicts about a third of the variability in individual storm properties, such as maximum intensity, showing their variability is dominated by synoptic conditions. Isolating the impact of present-day climate change on individual storms shows it contributes to about 0.1\% for storm-intensity variability, whereas their contribution to storms’ heat-anomaly variability is over three times greater, highlighting that focusing on variables directly tied to global warming offers a clearer attribution pathway.
\end{abstract}

\section*{Plain Language Summary}
Extratropical storms, which play a major role in shaping midlatitude weather, are influenced by both long-term climate patterns and short-term atmospheric variability. In this study, 84 years of atmospheric data combined with machine learning techniques are used to quantify the extent to which climate shapes storm activity. The results show that long-term averaged storm activity is strongly tied to the climate, whereas the behavior of individual storms, including their maximum intensity and track, is largely governed by short-term and small-scale variations. When measuring the impact of recent climate change on individual storms, the signal is extremely weak compared to the background of natural variability. In contrast, the climate change signal is much clearer for warm temperature anomalies associated with storms. These findings suggest that efforts to attribute specific midlatitude weather events to climate change should focus on factors that are more directly linked to long-term climate trends.

\section{Introduction}

When attempting to attribute extreme extratropical weather events, often associated with cyclones and anticyclones, to changes in Climatic Forcing (CF) \citep{Philip2022,Ginesta2023}, a major uncertainty arises from the Synoptic Variability (SV) of storms. Over extended periods, the climate is expected to be influenced primarily by CF, such as solar insolation, sea surface temperature anomalies, and the greenhouse effect. However, due to the chaotic nature of weather, the dynamics of individual storms exhibit large SV, which results from the fine structure of the meteorological conditions during their formation. Although SV and CF play significant roles in determining the intensity and trajectory of individual storms, their relative importance has yet to be quantified. Bridging this knowledge gap can uncover the importance of the dynamical response of storms to changes in CF when studying individual weather events. Therefore, in this study, using a statistical approach, we attempt to quantify the importance of CF to the growth of individual storms and the climatology of storm activity.

The effect of CF on midlatitude climate and weather has traditionally been studied by decomposing the flow into a mean (spatially or temporally averaged) and an eddy component. In this framework, the mean flow largely reflects climatological features such as the jet stream, while eddies describe synoptic-scale weather systems. Hence, given a particular CF, serving as a boundary condition, one expects a characteristic midlatitude flow pattern, consisting of a mean flow and a spectrum of eddies. Our central goal is to quantify how strongly these synoptic eddies (storms) depend on the mean flow, both in the time-averaged sense (climatological storm activity) and at the scale of individual storm growth.

The interaction between the mean flow and midlatitude storms is well studied, both from theoretical and idealized perspectives \citep{charney1947,Eady1949,Phillips1954,Harnik2004}, climatological alternation in the intensity of the storm tracks \citep{Nakamura1992,schemm2019efficiency}, and the effect on individual storms \citep{Orlanski1991,Orlanski1993,Riviere2006,Schemm2020,Hadas2023,hadas2025lagrangian}. Despite this extensive literature, the fundamental question—namely, the degree to which the mean flow dictates the characteristics of individual storms, has not been addressed quantitatively.

In this study, we use 84 years of ERA-5 reanalysis (Sec.~\ref{sec:ERA5}) and tracks of about 100,000 cyclones and 50,000 anticyclones (Sec.~\ref{subsec:Storm-tracking-1}) to quantify the extent to which storm activity is determined by the mean flow from a statistical perspective. To investigate this relationship, we use two modeling approaches. First, the climatological model (Sec.~\ref{sec:Unet}) tests whether the link between the mean flow and the climatological storm activity is unique (Sec.~\ref{sec:Eulerian}). Second, the single storm model (Sec.~\ref{subsec:method_single_storm}) evaluates how strongly the mean flow influences the properties of individual storms (Sec.~\ref{sec:Lagrangian}). Finally, we discuss the implications of our findings for the attribution of weather events and climatic changes to current anthropogenic climate change (Sec.~\ref{sec:CC}).  

\section{Methods}

\subsection{Reanalysis data} \label{sec:ERA5}

Data from the ERA-5 reanalysis by the European Center for Medium-Range Weather Forecasts \citep{Hersbach2020,Soci2024era5}, spanning from 1940 to 2023, is employed for current climate assessment. The ERA-5 reanalysis provides estimates of atmospheric variables with a horizontal resolution of 31 km and 137 vertical levels. Three-hourly Sea Level Pressure (SLP) data is utilized to track cyclones and anticyclones, while three-hourly temperature ($T$), specific humidity ($Q$), zonal wind ($U$), and meridional wind ($V$) at pressure levels 300, 500 and 850~hPa are utilized to estimate the atmospheric state and its climatology.

\subsection{Storm tracking\label{subsec:Storm-tracking-1}}

A feature point tracking algorithm \citep{hodges_1995,Tamarin2016a,Tamarin2017a} is applied to SLP data to identify and characterize extratropical cyclones and anticyclones. To reduce noise, the data is smoothed to a T63 resolution. The background, defined as zonal wavenumbers 0-4, is removed to isolate the synoptic scale dynamics. Only SLP anomalies deeper than 8 hPa are tracked to focus on significant events. After identification, systems are tracked, and their location and intensity are recorded if they persist for more than 48 hours and move more than 500 km westward. Additionally, systems that reach peak intensity over topography higher than 1 km are filtered out. 

The tracking algorithm returns the positions and depth (in hPa) of the storms from genesis to lysis. The output of the tracking algorithm is used to quantify the growth of storms using four metrics: the maximum intensity, defined as the maximum depth of the storm normalized by the sinus of latitude \citep{sanders1980synoptic}, the growth time, defined as the time between identification and maximum intensity, and the meridional and zonal rate of propagation of the storm from genesis to maximum intensity. In addition, the maximum 850~hPa temperature anomaly associated with a cyclone is defined as the maximum temperature anomaly around the cyclone center in a composite of $15^{\circ}\pm$ in the zonal direction and $10^{\circ}\pm$ in the meridional direction. An anomaly is defined as the deviation from the 84-year grid and time of the year average. 

\subsection{Climatological model}\label{sec:Unet}

The connection between the mean flow and climatological storm activity is treated as an image-to-image regression problem (Fig.~\ref{fig:Eulerian}), solved using Convolutional Neural Network (CNN),  due to its ability to efficiently learn spatial hierarchies and patterns in structured data. The model input is the three-dimensional distribution of $\overline{U}$, $\overline{V}$, $\overline{Q}$, and $\overline{T}$ over the midlatitudes (plus and minus 20$^\circ$-80$^\circ$ in for the NH and SH, respectively), where bar represent 90-day moving mean (45 days in each direction). The output is the distribution of vertically integrated Eddy Kinetic Energy:
\begin{align}
   \text{EKE}=\frac{1}{g}\int_{850}^{300}\frac{\overline{u'^{2}+v'^{2}}}{2}dp , \label{eq:EKE}
\end{align}
 where the bar is defined the same as above, the prime is defined as a deviation from a 10-day moving mean to focus on synoptic activity, and the integration is performed between 300 and 850~hPa. The EKE is calculated over the same domain. divided into train-validation-test set in ratios of $0.5-0.2-0.3$. An in-depth discussion of the data appears in Sec.~S1. 

Averaging the data over time substantially simplifies the regression task by reducing the effective image size in the frequency domain. Because atmospheric waves typically follow a dispersion relation—where slower waves tend to be larger—this temporal averaging further diminishes the impact of small-scale waves on variability (Fig.~S1 a-c). This simplification achieved can be quantified by the number of principal components required to capture 90\% of the variability in the dataset. Analyzing this reveals that filtering reduces the number of required components by 80\% (Fig.~S1d). In addition, time averaging significantly simplifies the dynamics, as it nullifies (or at least significantly reduces) the contribution of many physical processes to the overall system \citep{Vallis2017atmospheric}.  Consequently, much simpler models can simulate the relation between mean and eddy flow, relative to numerical weather prediction models \citep{Lam2023,Bi2023,Kochkov2024neural}.

The model is built out of a series of convolutional layers \citep{krizhevsky2012imagenet,guan2023learning}. Given an input, it predicts the shape and scale parameters of the underlying Gamma distribution, which are optimized using negative log-likelihood loss (Eq.~2 in the supporting information), similar to \cite{guillaumin2021stochastic}. An in-depth discussion of the model, including the loss function, structure, hyperparameters, and steps taken to avoid overfitting, is provided in Sec.~S2.

\subsection{Single storm model} \label{subsec:method_single_storm}

The extent to which the mean flow sets the growth of individual storms is tested by constructing a model that predicts the average outcome of a storm's growth based on the mean flow it experiences at genesis. The model input is the three-dimensional composites of $\overline{U}$, $\overline{V}$, $\overline{Q}$, and $\overline{T}$ around the storm at genesis. The model output is the outcome of growth, namely the maximum intensity, the growth time, and the meridional and zonal rate of propagation of the storm from genesis to maximum intensity (Sec.~\ref{subsec:Storm-tracking-1}). Tracking all cyclones and anticyclones between the years 1940 and 2023 results in approximately 100,000 cyclones and 50,000 anticyclones. An in-depth discussion of the data appears in Sec.~S3.

The model consists of multiple convolutional layers, each followed by max-pooling to reduce dimensionality, thus performing feature extraction. The resulting features then pass through two dense layers to regress on the storm properties. To assess the model’s robustness, cross-validation is performed by reshuffling the train–validation–test splits and training/testing the model 50 times. Additionally, we compare the CNN to a dense neural network (no feature extraction) and a random forest \citep{breiman2001random}, implemented via \texttt{RandomForestRegressor} and \texttt{MiniBatchKMeans} from scikit-learn \citep{Scikit-learn}. Further details on the model architecture, hyperparameters, and this comparison can be found in Sec.~S4.

\section{Results}

\subsection{To what extent does the mean flow set the climatological storm activity?}\label{sec:Eulerian}

Climatological storm activity becomes largely independent of synoptic variability (SV) when averaged over sufficiently long timescales. Consequently, if the mean atmospheric state captures most of the climate forcing (CF) information, it should explain much of the variability in climatological storm activity. To test this, we evaluate how well a climatological model can predict EKE maps using only mean-flow fields (Sec.~\ref{sec:Unet}).

\begin{figure}
\begin{centering}
\includegraphics[width=\textwidth]{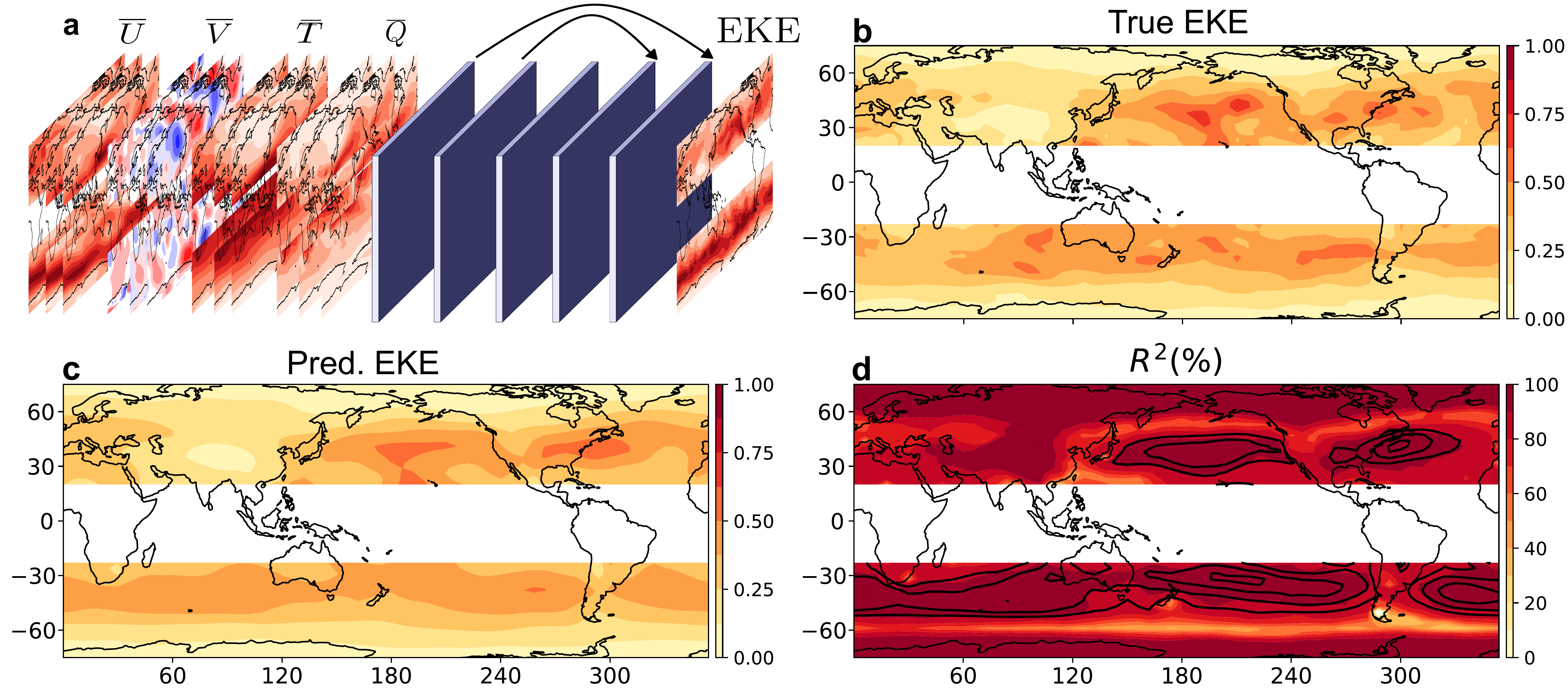}
\par\end{centering}
\caption{\label{fig:Eulerian}(a) Illustration of the climatological model. (b,c) True and predicted EKE (MJ~m$^-2$,  computed using Eq.~\ref{eq:EKE}). (d) $R^2$ score, calculated over the test set (shading), and the multi-annual mean EKE, depicted as contours ranging from 0.36 to 7.6 (MJ~m$^-2$) in intervals of 0.04.}
\end{figure}

Comparing the model’s predictions with the climatological EKE averaged from September 28 to December 27, 2017 (Fig.~\ref{fig:Eulerian}b vs. Fig.~\ref{fig:Eulerian}c) shows that the model successfully reproduces major features such as the EKE “hot spots” along the Northern Hemisphere Atlantic and Pacific storm tracks, as well as many smaller-scale seasonal and interannual variations. To quantify performance, we use the coefficient of determination:
\begin{equation}
R^{2}=1-\frac{\sum\left(y_{i}-y_{\text{pred}}\right)^{2}}{\sum\left(y_{i}-\overline{y}\right)^{2}},\label{eq:R2}
\end{equation} 
where $y_{i}$ and $y_{\text{pred}}$ are the true and predicted EKE at each grid point over time (test set), and $\overline{y}$ is the overall mean EKE. As shown in Fig.~\ref{fig:Eulerian}d, the model achieves $R^2$ values above 90\% in most regions (shading), especially in regions of high climatological storm activity (contours). Overall, about 93\% of the total EKE variability is explained on average. These results confirm that the mean flow contains the core CF information essential for determining the climatology of storm activity.

Shortening the averaging window for EKE reduces the model’s predictive skill. However, as long as the window remains longer than the synoptic scale (roughly 10 days), the model still explains above 50\% of the variability (Fig.~S3). Hence, the mean flow dominates storm-activity variability up to the scale of a single synoptic cycle. Moreover, model performance is relatively insensitive to which atmospheric variables are provided (Fig.~S4), with a single variable alone accounting for over 90\% of the variance. This finding reflects that temporal averaging enforces consistency among mean flow variables. For the dry variables, the thermal wind balance quantitatively demonstrates this, as it allows the inference of the mean atmospheric wind profile based on the mean temperature profile and vice versa. The consistency with the mean atmospheric water content can be understood qualitatively since atmospheric dynamics influence the injection, removal, and distribution of water vapor, which couples back through diabatic effects, and therefore, their climatologies are tightly connected.

\subsection{To what extent does the mean flow set the growth of individual storms?\label{sec:Lagrangian}}

While the mean flow largely sets the climatology of storm activity, individual storms can also be influenced by SV. To quantify the relative importance of the mean flow, we construct a model that predicts storm growth based solely on the mean flow $\overline{X}$ (Sec.~\ref{subsec:method_single_storm}). Because $\overline{X}$ encapsulates most of the relevant CF information for storm climatology (Sec.~\ref{sec:Eulerian}), it is expected to provide an effective representation of CF’s influence on individual storm dynamics. Meanwhile, SV is captured by the remainder of the atmospheric initial state, $X^{'}=X-\overline{X}$, where $X$ is the full initial state. Since storm dynamics also depend on $X^{'}$, there is a range of possible growth outcomes for any given $\overline{X}$ \citep{Mana2014}. Consequently, a model that relies only on $\overline{X}$, cannot distinguish among these different outcomes. Instead, with a mean square error loss, it learns the single best-fit prediction for the observed distribution of outcomes, i.e., the mean of that distribution \citep{Delsole2022analysis}. If $\overline{X}$ strongly drives storm growth variability, each $\overline{X}_i$ leads to a relatively narrow distribution of possible outcomes, separated from other $\overline{X}_{j\ne i}$, yielding a lower model error. Conversely, if SV dominates, each $\overline{X}$ is associated with a broad distribution of outcomes, which diminishes the predictive skill of the mean-flow-only model, resulting in higher model error.

\begin{figure}
\begin{centering}
\includegraphics[width=0.9\textwidth]{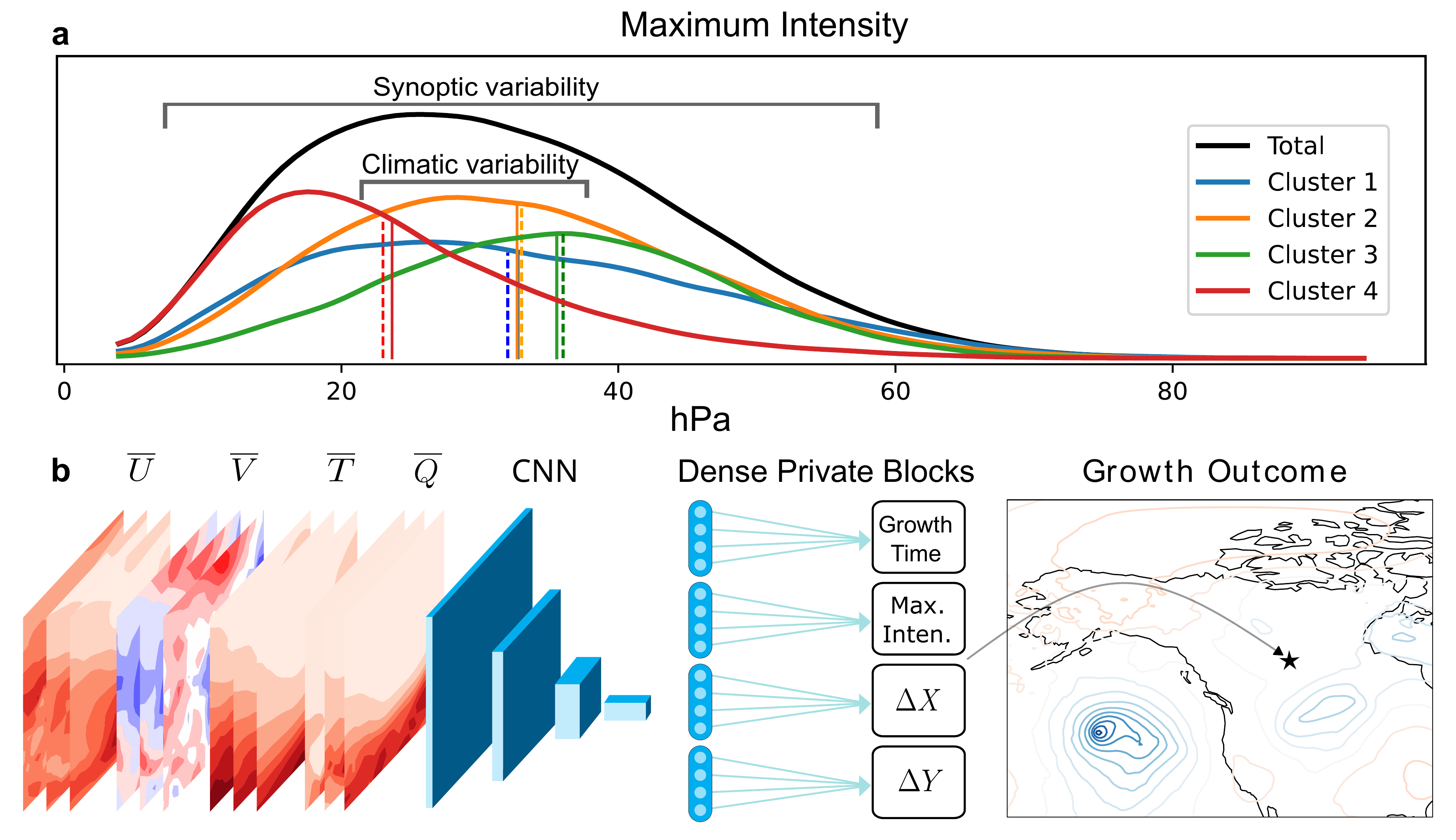}
\par\end{centering}
\caption{\label{fig:Lagrangian}(a) The total distribution of storms' maximum intensity (hPa, black curves), and the distribution for four characteristic mean flow conditions identified using K-mean clustering. Solid and dashed vertical lines represent the mean of the distribution and the model mean prediction, respectively. The intensity distribution curves are smoothed using a cubic Hermite spline. (b) Illustration of the single storm model: the model gets the three-dimensional mean flow at genesis in a composite box centered around the storm, paths it through a CNN for feature extraction, and predicts the outcome of the storm's growth based on the extracted features.}
\end{figure}

To illustrate this concept, we approximate the distribution of maximum storm intensities under different mean flow conditions by clustering storms into four groups using K-means applied to $\overline{X}$ (Fig.~\ref{fig:Lagrangian}a). The resulting distributions are distinct, indicating that even coarse sorting by mean flow yields differences in growth outcomes. However, the within-cluster spread, which is approximately the SV, is large relative to the differences between the cluster means, which is due to the variability of CF. To go beyond this qualitative demonstration, we train a neural network to learn the mapping between $\overline{X}$ and the expected storm growth (Fig.~\ref{fig:Lagrangian}b, see Sec.~\ref{subsec:method_single_storm} for details). The network’s predictions closely match the average maximum intensity observed within each cluster (Fig.~\ref{fig:Lagrangian}a, solid vs. dashed vertical lines). The model’s mean square error, normalized by the total variance in storm intensity, then provides a quantitative estimate of the contribution of SV to storm growth variability. One minus this value gives the contribution of climatic variability, and also corresponds to the model's $R^2$.

\begin{figure*}
\begin{centering}
\includegraphics[width=0.9\textwidth]{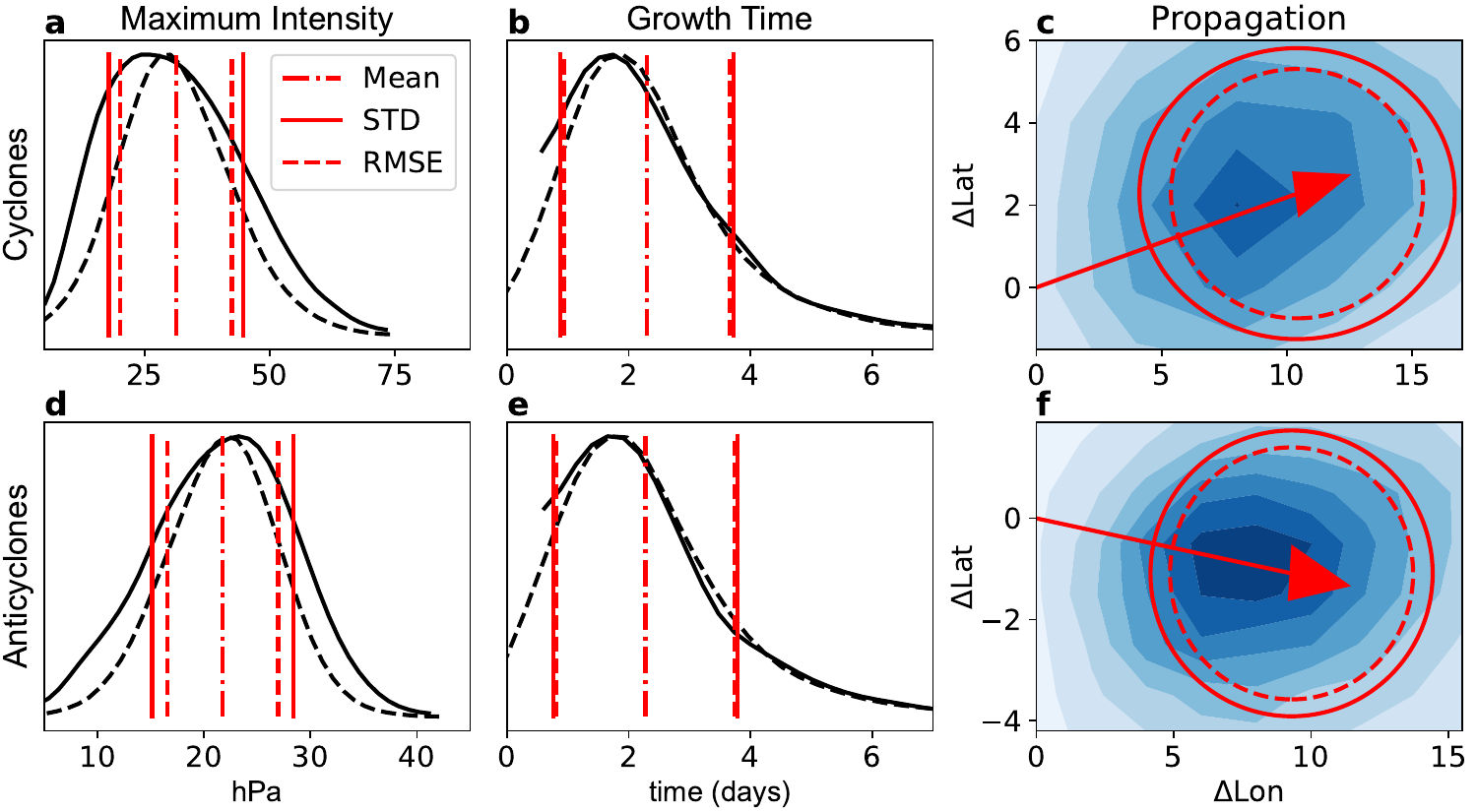}
\par\end{centering}
\caption{\label{fig:Lagrangian_fig}(a) Distribution of cyclones' maximum intensity (hPa, black curve), overlaid with the distribution of the model residual ($y-y_{\text{pred}}$, black dashed). The red dotted-solid line marks the mean, the red solid line marks the standard deviation (STD) of the distribution, and the red dashed line marks the model root mean square error (RMSE), which is the standard deviation of the dashed residual curve. The residual distribution is shifted so that it has the same mean as the overall distribution. (b) Same as (a) but for growth time (days). (c) Two-dimensional distribution of the zonal and meridional rate of propagation (degrees day$^{-1}$) from genesis to maximum intensity (color, abscissa, and ordinate, respectively), along with the average propagation vector (red arrow), the standard deviation (red solid ellipse), and the model RMSE (red dashed ellipse). The ellipse axes show the standard deviation (RMSE) of the data (model) along each axis. (d-f) Same as (a-c) but for anticyclones.} \end{figure*}

Fig.~\ref{fig:Lagrangian_fig} compares the single-storm model's residual distribution against the overall distribution to illustrate the relative contributions of mean flow and SV. As indicated by the similar widths of the residual (dashed lines) and the original distribution (solid lines), most of the variability in individual storm properties cannot be explained by the mean flow. Quantitatively, for cyclones (anticyclones), the climatic variability accounts for only about 30\% (38\%) of the the maximum intensity variability (Fig.~\ref{fig:Lagrangian_fig}a,d), 7\% (7\%) of the growth-time variability (Fig.~\ref{fig:Lagrangian_fig}b,e), 34\%(25\%) of the zonal propagation rate variability, and 25\% (23\%) of the meridional propagation rate variability (Fig.~\ref{fig:Lagrangian_fig}c,f). The low climatic variability values show that most variability among storms on Earth arises from SV rather than from CF.

A series of sensitivity tests confirms the robustness of our single-storm model results. First, using only a Dense Neural Network (DNN) or a nonparametric Random Forest regressor yields nearly identical performance (Table~S1). Second, training the model on at least 20,000 samples appears sufficient for convergence (Fig.~S5). Finally, the model’s predictive skill increases significantly only when the averaging window $\overline{X}$ is shorter than about 10 days—an interval that begins to capture more synoptic variability (Fig.~S6).

\subsection{To what extent does climate change affect the growth of individual storms?}\label{sec:CC}

Due to anthropogenic forcing, the mean climate has changed in the last decades \cite{Vallis2015,Forster2021,Woollings2023}. In this section, the portion of single-storm variability attributable to climate change is isolated. This is achieved by comparing the predictability found in Sec.~\ref{sec:Lagrangian} to a modified test set in which the climate trend is “scrambled,” enabling a distinction between the long-term climate signal and other climate-driven variability. Following prior studies \citep[e.g.,][]{Woollings2023}, the climate trend at each grid box is determined by linearly regressing mean flow on the year for each date. A modified test set is then created by removing the correct trend and adding the trend from a random year, thus preserving seasonal and interannual variability while eliminating the actual long-term signal. Finally, the storm model’s performance (Sec.~\ref{subsec:method_single_storm}) on this modified dataset is compared with that on the original test set, and the drop in skill is attributed to the climate change trend. 

\begin{figure*}
\begin{centering}
\includegraphics[width=0.45\textwidth]{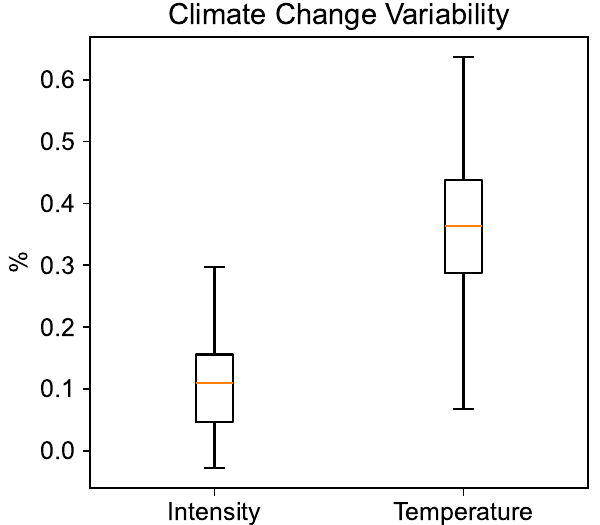}
\par\end{centering}
\caption{\label{fig:LagrangianCC}Box-and-whisker plots of the contribution of climate change to variability (\%) for maximum intensity (Intensity) and temperature anomaly (Temperature). The box spans the interquartile range (IQR, Q3 minus Q1), the whiskers extend to the most extreme data points within 1.5 IQR, and the line inside the box marks the median.}
\end{figure*}

In this section, only the years 1980-2023 were used, in order to focus on a period where the climate trend is large (as can be assessed qualitatively from the global mean surface temperature trend, \cite{Forster2021}), and to reduce biases related to data availability \citep{Soci2024era5}. To estimate the uncertainty, we train and test the model 50 times, where each time we assign different storms for training, validation, and testing. From this comparison, the climate trend in the mean flow explains approximately $0.10(\pm0.01)\%$ of cyclone intensity variability (Fig.~\ref{fig:LagrangianCC}). Thus, accurately discerning such a weak signal poses a substantial challenge for attribution efforts, especially given the dominance of synoptic variability found in Sec.~\ref{sec:Lagrangian}.

 Nevertheless, these results do not imply that climate change is undetectable in the midlatitudes. As shown by many studies and supported here (Fig.~\ref{fig:Eulerian}), temporally averaging storm activity diminishes the role of synoptic variability. Once the long-term signal rises above natural variability, its detection becomes feasible \citep{Priestley2022,Chemke2022,chemke2024human,Shaw2024emerging}. Furthermore, certain event attributes respond more directly to climate change, making them more amenable to attribution \citep{Trenberth2015,catto2019future}. For example, temperature anomalies accompanying cyclones (see Sec.~\ref{subsec:Storm-tracking-1}) are expected to show a stronger climate-change signal due to robust warming. Indeed, re-training the model to predict temperature anomalies and scrambling their trend shows that climate change currently contributes to about $0.35(\pm0.017)\%$ of the variability (Fig.~\ref{fig:LagrangianCC}), more than three times higher than for cyclone intensity. These findings underscore that, while attributing individual storm intensity changes to climate forcing remains difficult, examining variables more directly linked to a warming climate can offer a clearer attribution pathway.

\section*{Open Research Section}
No new data sets were generated during the current study. The three-dimensional atmospheric variables on pressure levels (e.g., zonal and meridional wind) are available from Copernicus through the dataset "ERA-5 hourly data on pressure levels from 1940 to present" \citep{Hersbach2023PressureLevels}. Sea level pressure data is available from Copernicus through the dataset "ERA-5 hourly data on single levels from 1940 to present" \citep{Hersbach2023SingleLevels}.

\section*{Conflict of Interest Statement}

The authors have no conflicts of interest to disclose.

\acknowledgments
This research has been supported by the Azrieli fellowship and the Israeli Science Foundation (Grant 996/20). We Thank Ido Aizenbud, Jonathan Kavitzki, Dotan Gazith, Gidi Yoffe, and Nimrod Gavriel for the meaningful discussion.

\bibliography{orsbib}

\section*{Supplementary Information}

\noindent\textbf{Text S1: Data for the climatological model}
The resolution of both the input and output images is 6$^{\circ}$ in the zonal direction, 3$^{\circ}$ in the meridional direction, and the levels 300, 500, and 850~hPa are used in the vertical. Using this resolution, the inputs and outputs are images of 20 by 60 pixels, with 9 channels for the input (three levels of three variables) and 3 channels for the output (three levels of one variable). The meridional direction of the SH data is flipped to match the NH.

The data is sampled in 10-day intervals to ensure significant differences between data points while obtaining a large dataset. Overall, there are about 6,000 data points (36 images per year for each hemisphere for 84 years) divided into train-validation-test set in ratios of $0.5-0.2-0.3$. The division between sets is done by splitting the data into chunks of 108 consecutive samples (i.e., about 3 years), with a margin of five samples between chunks that were not included in any data sets. Then, given that the maximum window that was used was 90 days (45 on each side), no data leak was introduced. 

For training and validation, data augmentation is performed by randomly shifting the central longitude pixel, leveraging the system's symmetry with respect to the central longitude. Each epoch consists of three repetitions of the input data, with each sample shifted by a different randomly chosen integer.

The input images are normalized using the mean and standard deviation computed along the zonal direction and over time, ensuring symmetry is preserved in these dimensions. The output is normalized using the global standard deviation (computed across all pixels and time) to prevent feeding extremely large or small values into the model, improving numerical stability. However, the mean is not removed to maintain the positivity of the output. Notably, the loss function itself accounts for per-pixel normalization based on the input (see next section).

\noindent\textbf{Text S2: The climatological model}

To capture the variability in each pixel’s EKE, which can differ substantially based on the mean flow, we adopt a negative log-likelihood framework that estimates variance on a per-pixel basis. Since EKE is strictly positive, we model its distribution using a Gamma distribution:
\begin{align}
p(y; \alpha, \beta) = \frac{y^{\alpha - 1} e^{-y/\beta}}{\beta^\alpha \, \Gamma(\alpha)}, \quad y > 0, \quad \alpha > 0, \quad \beta > 0.
\end{align}
where $\alpha$ (shape) and $\beta$ (scale) parameters define the distribution and $\Gamma$ is the Gamma function. The mean and standard deviation of this distribution are $\alpha\beta$ and $\alpha\beta^2$, respectively. 

Following \cite{guillaumin2021stochastic}—but substituting the Gamma PDF—the negative log-likelihood (loss) function we optimize is:
\begin{align}
L(y;\alpha(X),\beta(X)) = \sum_{i=1}^{N \text{ pixels}} \left[ (\alpha_i-1)\ln y_i - \frac{y_i}{\beta_i} - \alpha_i \ln \beta_i - \ln \Gamma(\alpha_i) \right].
\end{align}
where $y_i$ is the true EKE in pixel $i$ and $\alpha_i, \beta_i$ are predicted from the input. Hence, the model outputs two images per input image: one for the shape parameters and one for the scale parameters, enabling joint optimization of both moments. To compute the  $R^2$ score, we set the model’s EKE prediction to $y_{\text{pred}}=\alpha\beta$. Empirically, this approach achieves about 7\% higher $R^2$ (92\% vs. 85\%) than a model trained using mean square error.

Our model is built from convolutional layers, each followed by a ReLU activation function. To address the anisotropic input dimensions, we use kernels of size 3 $\times$ 9 (meridional $\times$ zonal). We apply zero-padding in the meridional direction and circular padding in the zonal direction to preserve image dimensions. Each layer has eight filters, effectively capturing spatial features. The final layer has two filters—one for the shape parameter and one for the scale parameter—and uses a softplus activation ($\log{1+e^x}$) to ensure positivity.

Systematic tuning of the number of layers and filters (Fig.~S2) shows that performance increases with the total number of parameters up to around 10$^4$. This improvement behaves similarly whether we add more layers or more filters. Consequently, we use five layers with eight filters each, yielding about 10$^4$  parameters and an $R^2$ above 90\%.

To avoid overfitting the model parameters, we monitor the validation loss after each epoch (i.e., after each pass over the training set), selecting the model checkpoint with the lowest validation loss. To prevent overfitting to hyperparameters, all final results reported in the paper are evaluated on a separate test set, which is never used during training or hyperparameter tuning.

Using default settings in Scikit-Learn, LinearRegression, MLPRegressor, and RandomForestRegressor achieve $R^2$ scores of 86\%, 90\%, and 90\%, respectively. Although these simpler models perform reasonably well, our CNN-based approach outperforms them. Furthermore, given the high dimensionality of both inputs and outputs, the CNN model runs and tunes significantly faster, making it more practical for this application.

\noindent\textbf{Text S3: Data for single storm model}

The model input is the three-dimensional composites of the mean flow, which extend $\pm60^{\circ}$ in the zonal direction and $\pm30^{\circ}$ in the meridional direction, with resolution of 6$^{\circ}$ in the zonal direction and 3$^{\circ}$ in the meridional direction. This relatively low resolution is chosen as the mean flow is associated with large-scale phenomena, and therefore increasing the resolution would not change the results significantly. Therefore, the overall shape of the input is 21 by 21 pixels with 9 channels (three variables times three vertical levels). The model output is the outcome of growth, namely the maximum intensity, the growth time, and the meridional and zonal propagation of the storm from genesis to maximum intensity. The storms' properties in the input and output are based on tracks of storms obtained through a feature tracking algorithm. Tracking all cyclones and anticyclones between the years 1940 and 2023 results in approximately 100,000 cyclones and 50,000 anticyclones, which are divided between train-validation-test in a ratio of $0.6-0.2-0.2$. The input images are normalized by the grid-wise average and standard deviation. The input and output vectors are also normalized by the mean and standard deviation. 

\noindent\textbf{Text S4: The single storm model}

The first part of the model processes the image input to perform feature extraction. The basic building blocks of this part are convolutional blocks, which transform the input using two-dimensional convolution, followed by Relu activation and Max Pooling with a kernel of 2-by-2. Three of the above layers are stacked. The first convolutional layer has 4 filters, and each time a pooling is performed, the number of filters in the next layers is doubled. 

The second part of the model is built out of dense layers. The output of the CNN is passed to a private dense network for each predicted property (four in total), which is built out of two dense layers with 8 nodes, followed by Relu activation. The outer layer is built out of a dense layer with a single node and linear activation, which allows regression. 

The performance of the model is assessed using a mean square error loss function. The mean square error is also used to assess the SV.

 Shortening the averaging window (Fig.~\ref{fig:Lagrangian_R2_window}) significantly affects the model performance only when the averaging time is shorter than the synoptic time scale, as it includes knowledge about the SV. Training the model on part of the storms in the training set shows that about 10,000-20,000 storms are required to achieve convergence (Fig.~\ref{fig:Lagrangian_data}).

The CNN’s performance is compared to two baseline models: a four-layer Deep Neural Network with 16 nodes per hidden layer,  and a Random Forest regressor in Scikit-Learn with 250 estimators, bootstrap enabled, a maximum depth of 30, a maximum feature fraction of 0.05, a minimum of 16 samples per leaf, and a minimum of 20 samples per split. The results of the validation appear in Tab.\ref{tab:performance_val} and show that the CNN, DNN, and Random Forest perform very similarly, whereas the K-Means approach while improving with more clusters, remains below the performance of the other methods even with 100 clusters.

\clearpage
\renewcommand{\thefigure}{S\arabic{figure}}
\setcounter{figure}{0}
\begin{figure}[b]
\begin{centering}
\includegraphics[width=\textwidth]{ 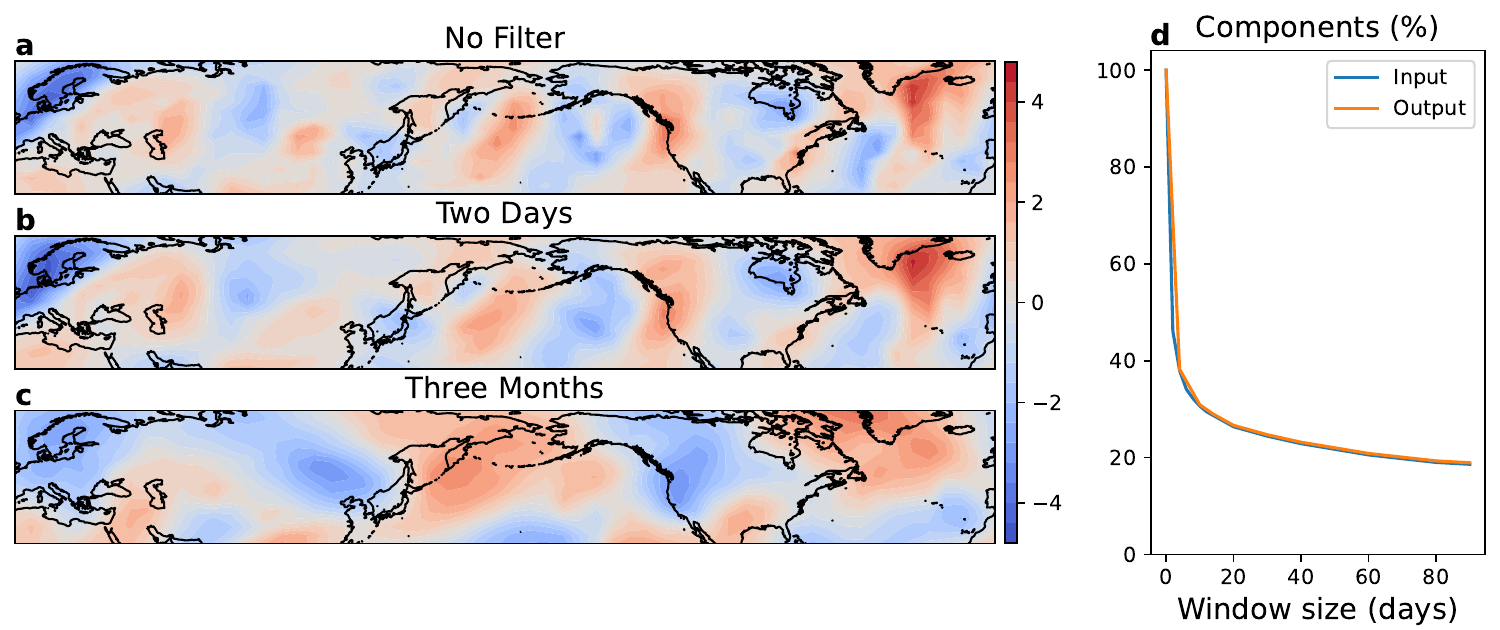}
\par\end{centering}
\caption{\label{fig:pca} (a-c) Snapshots of 300~hPa meridional wind normalized by the image standard deviation (unitless) for data that was not averaged, data which was averaged over two weeks, and data which was averaged over three months. (d) The percentage of principal components to explain 90\% of the variability in the 300~hPa wind. The percentage is defined as the number of principal components required relative to the data without filter, which required 255 components (10\% of the number of features in the original image).}
\end{figure}

\begin{figure}[b]
\begin{centering}
\includegraphics[width=\textwidth]{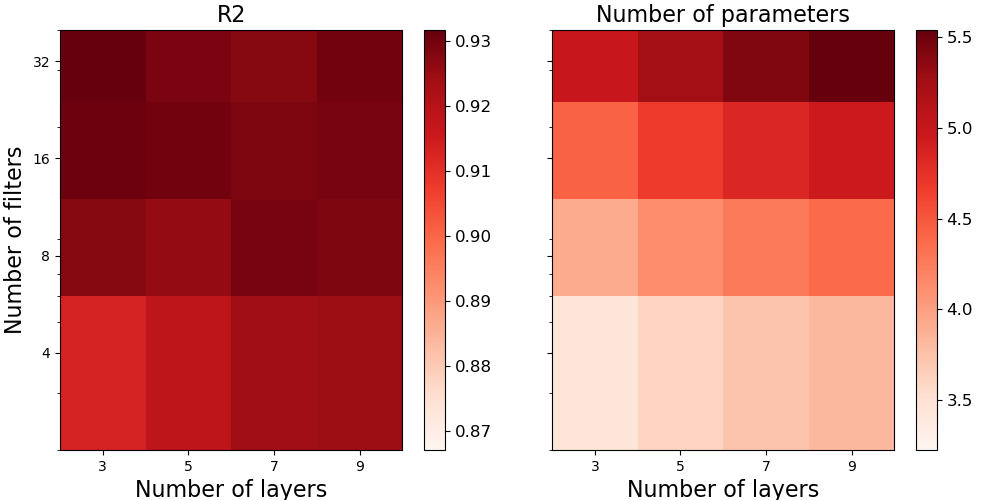}
\par\end{centering}
\caption{\label{fig:pca} The optimal $R^2$ calculated over the validation set (a) and the log10 of the number of parameters of the model as function of the number of filters and number of blocks.}
\end{figure}

\begin{figure}[b]
\begin{centering}
\includegraphics[width=0.5\textwidth]{ 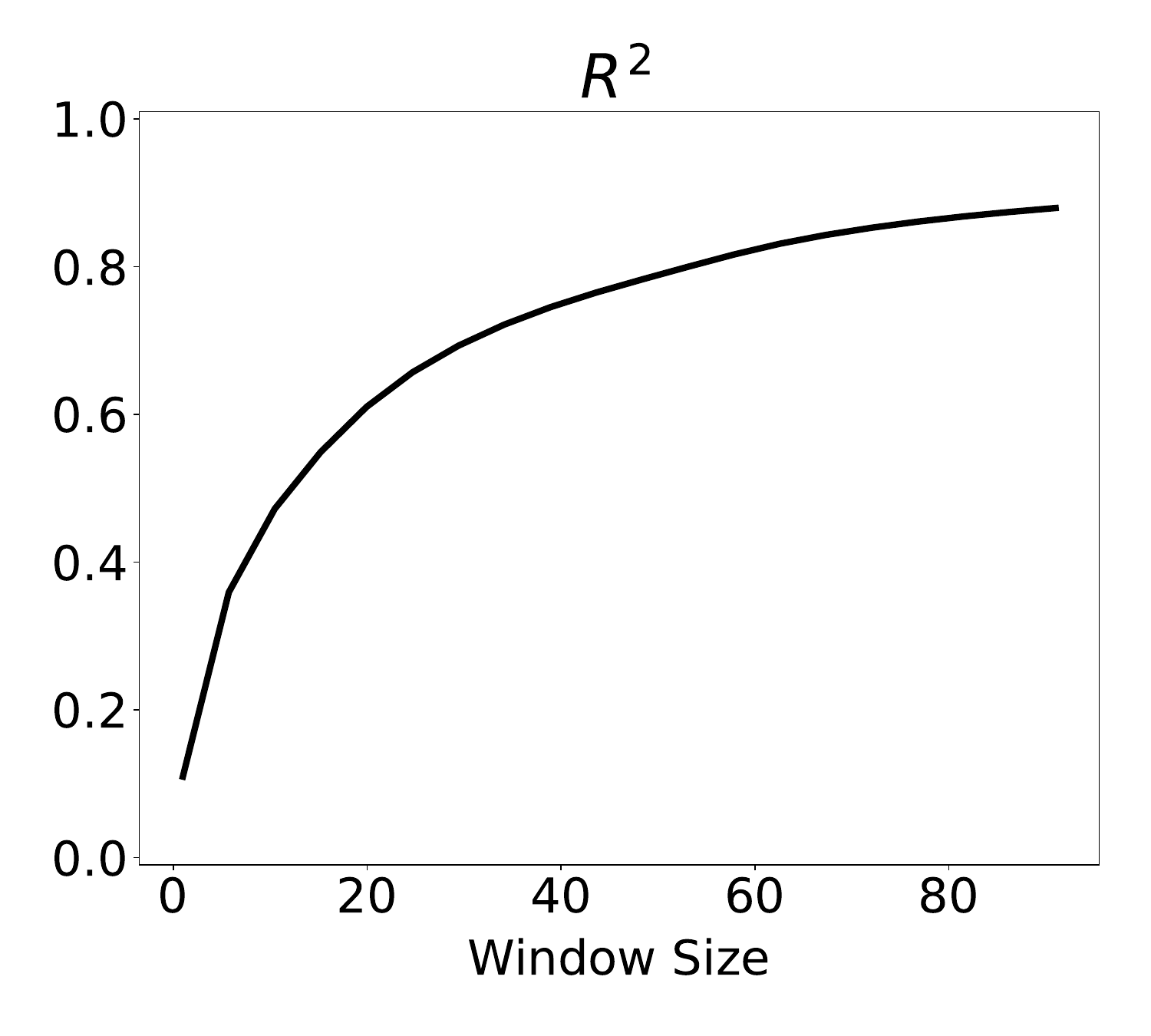}
\par\end{centering}
\caption{\label{fig:Eulerian_R2_window} The global mean $R^2$ value of the climatological model as a function of the EKE averaging window~(days). The data was smoothed using a Cubic Hermite spline.}
\end{figure}

\begin{figure}[b]
\begin{centering}
\includegraphics[width=0.5\textwidth]{ 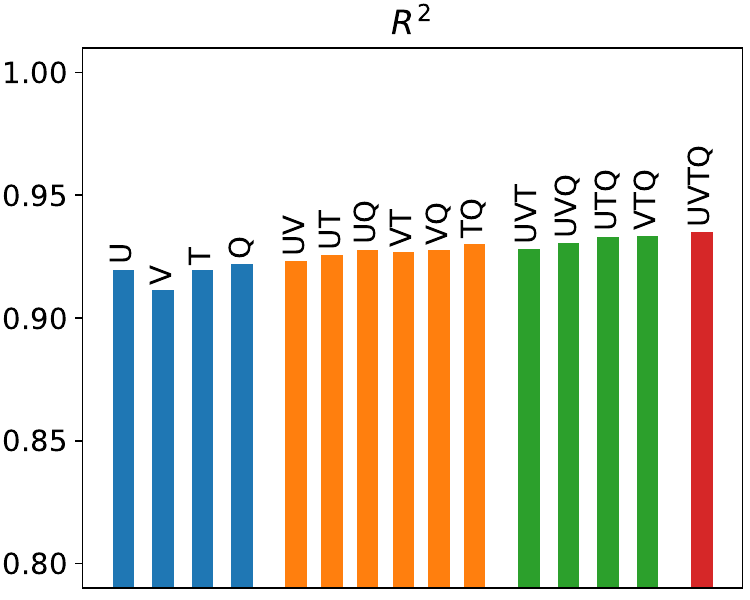}
\par\end{centering}
\caption{\label{fig:Eulerian_R2_vars} The global mean $R^2$ value of the climatological model as a function of the input variables, where $U$ is the zonal wind, $V$ is the meridional wind, $T$ is the temperature and $Q$ is the specific humidity.}
\end{figure}

\begin{figure}[b]
\begin{centering}
\includegraphics[width=0.5\textwidth]{ 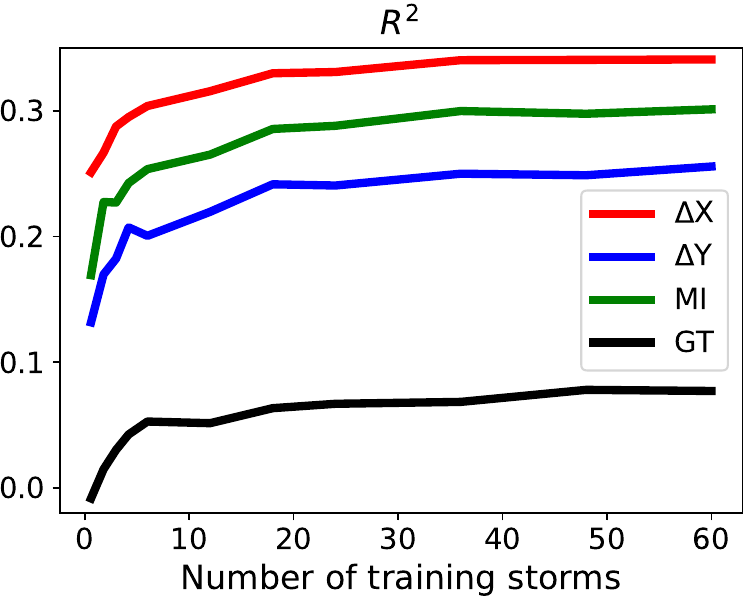}
\par\end{centering}
\caption{\label{fig:Lagrangian_data} The  $R^2$ value of the single storm model as a function of the number of storms used for training~(${\times}10^4$) for the Maximum Intensity (MI, red), Growth Time (GT, blue), meridional propagation ($\Delta$Lat., green) and zonal propagation ($\Delta$Lon., black).}
\end{figure}

\begin{figure}[b]
\begin{centering}
\includegraphics[width=0.5\textwidth]{ 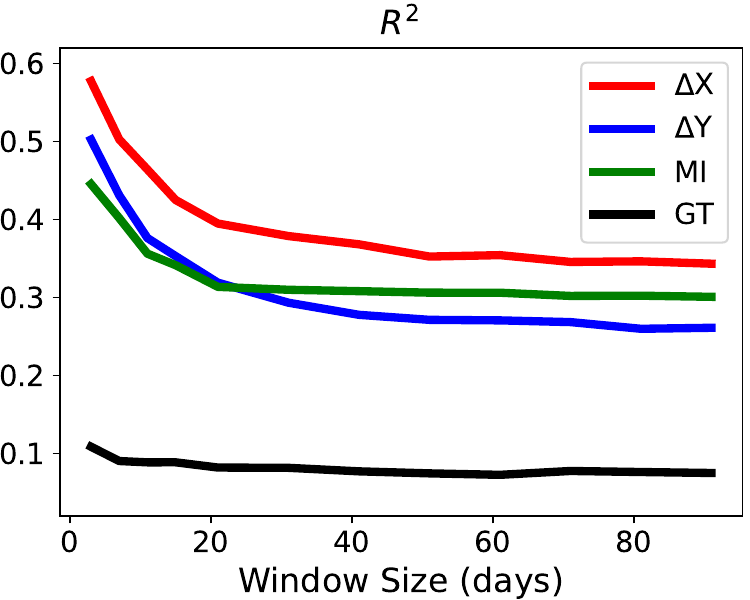}
\par\end{centering}
\caption{\label{fig:Lagrangian_R2_window} The  $R^2$ value of the single storm model as a function of the mean flow averaging period~(days) for the Maximum Intensity (MI, red), Growth Time (GT, blue), meridional propagation (${\Delta}$Lat., green) and zonal propagation (${\Delta}$Lon., black).}
\end{figure}

\begin{table}[]
    \centering
\begin{tabular}{l|cccc}
 & $\Delta$X & $\Delta$Y & $\Delta$ MI & $\Delta$ GT \\
\hline
CNN & 34 & 25 & 30 & 7 \\
DNN & 35 & 26 & 30 & 7 \\
RF & 36 & 27 & 31 & 8 \\
\end{tabular}
    \caption{$R^2~(\%)$ of the CNN, Dense Neural Network (DNN), Random Forest Regressor (RF) for prediction of the rate of propagation in the zonal and Meridional directions ($\Delta$ Lon. and $\Delta$ Lat.), the Maximum Intensity (MI) and the growth time (GT).}
    \label{tab:performance_val}
\end{table}

\end{document}